\documentclass[final,5p,times,onecolumn,unknownkeysallowed]{elsarticle}

\usepackage{doi}
\usepackage{hyperref}
\hypersetup{
	colorlinks=true,        
	linkcolor=blue,         
	citecolor=cyan,         
}

\usepackage[utf8x]{inputenc}
\DeclareUnicodeCharacter{2212}{\textendash}
\usepackage{graphicx}
\usepackage{amssymb}
\usepackage{amsmath}
\usepackage{amsthm}
\usepackage{lineno}
\usepackage{hyperref}
\usepackage{multirow}
\usepackage{bigints}

\usepackage{dcolumn}
\usepackage{bm}
\usepackage{color}

\hypersetup{colorlinks=true,linkcolor=cyan,citecolor=cyan,urlcolor=blue,bookmarks=true}\modulolinenumbers[200]

\biboptions{comma,sort&compress}

\newcommand{\be}{\begin{equation}}
\newcommand{\ee}{\end{equation}}
\newcommand{\ba}{\begin{eqnarray}}
\newcommand{\ea}{\end{eqnarray}}
\newcommand{\bal}{\begin{align}}
\newcommand{\eal}{\end{align}}

\newcommand{\bw}{\begin{widetext}}
\newcommand{\ew}{\end{widetext}}

\journal{Journal Name}

\begin{document}

\begin{frontmatter}

\title{\Large Weak cosmic censorship conjecture for the (2+1)-dimensional charged BTZ black hole in the Einstein-Gauss-Bonnet Gravity}

\author[mainaddress6]
{Ayyesha K. Ahmed}
\ead{aahmed.phdmath17sns@student.nust.edu.pk}
\author[mainaddress5,mainaddress2,mainaddress1,mainaddress4,mainaddress3,mainaddress7]
{Sanjar Shaymatov\cortext[cor2]{Corresponding author}\corref{cor2}}
\ead{sanjar@astrin.uz}
\author[mainaddress1,mainaddress3,mainaddress4]
{Bobomurat Ahmedov}
\ead{ahmedov@astrin.uz}

\address[mainaddress6]{Department of Mathematics, School of Natural Sciences (SNS), National University
of Sciences and Technology (NUST), H-12, Islamabad, Pakistan}
\address[mainaddress5]{Institute for Theoretical Physics and Cosmology, Zheijiang University of Technology, Hangzhou 310023, China}
\address[mainaddress2]{Akfa University,  Milliy Bog Street 264, Tashkent 111221, Uzbekistan}
\address[mainaddress1]{Ulugh Beg Astronomical Institute, Astronomy St. 33, Tashkent 100052, Uzbekistan}
\address[mainaddress4]{National University of Uzbekistan, Tashkent 100174, Uzbekistan}
\address[mainaddress3]{National Research University TIIAME, Kori Niyoziy 39, Tashkent 100000, Uzbekistan}
\address[mainaddress7]{Power Engineering Faculty, Tashkent State Technical University, Tashkent 100095, Uzbekistan}

\date{Received: date / Accepted: date}

\begin{abstract} 
It is well known that ($2+1$) dimensional charged BTZ (Banados, Teitelboim, Zanelli) black hole can be overcharged by a charged scalar field and a charged particle in contrast to their analogues in ($3+1$) and higher dimensions. In this regard, it may play an important role to understand more deeply the properties of the ($2+1$)-dimensional charged black hole in the Einstein-Gauss-Bonnet (EGB) gravity. In this paper, we test the validity of the weak cosmic censorship conjecture for the ($2+1$)-dimensional charged black hole in novel EGB theory derived recently by Henniger et. al (2021). We show that the minimum energy that particle can have at the horizon becomes negative for both an extremal and nearly-extremal BTZ black holes in EGB gravity. It is proven  that both extremal and nearly-extremal $(2+1)$ dimensional BTZ black hole could be overcharged in EGB theory, leading to the violation of the {weak cosmic censorship conjecture (WCCC)}, which is in good agreement with the results obtained for the classical BTZ black hole.
 
\end{abstract}


\end{frontmatter}

\linenumbers

\section{Introduction}

By the detection of gravitational waves from coalescence of black holes in close binary systems~\cite{Abbott16a,Abbott16b} and observation of supermassive black hole M87  shadow~\cite{Akiyama19L1,Akiyama19L6}, it has become quite easy to understand the nature of the spacetime geometry and the phenomenon of gravitational interaction in the strong field regime. In spite of this fact, {Einstein theory of gravity} is restricted e.g. due to its non-applicability for the physical singularities appeared. It was recently proposed {that} the novel Einstein-Gauss-Bonnet (EGB) theory could exist even in $D=3,4$ dimensions as viable alternatives to black holes in general relativity (GR), avoiding the Lovelock's theorem by rescaling the Gauss-Bonnet (GB) coupling constant (i.e., the limit $D\rightarrow 4$ ($D\rightarrow 3$ ))~\cite{Glavan20prl}. For this   
theorem~\cite{Lovelock71} the cosmological constant would appear only within the GR in dimension $<5$.

It is worth noting that, in contrast to the {Einstein theory of gravity}, the causal structure in $4D$ EGB gravity differs from its counterpart, i.e.,  the region around singularity is time-like while it behaves space-like for {Einstein theory of gravity}~\cite{Dadhich20egb}. The properties of the 3D and 4D-EGB black holes have attracted strong attention in recent years regardless of the fact that the validity of the EGB theory has come under severe scrutiny. {The main arguments against the EGB theory are follows: (\textit{i}) the issue on the redefinition of the GB term related to the validity of its rescaling that
would be possible only for systems with certain symmetries~\cite{Hennigar20egb,Arrechea20egb}, (\textit{ii}) the one related to the action, that is EGB theory has no a nontrivial four-dimensional limit and thus it may remain in dimension $>4$~\cite{Gurses20egb,Mahapatra20egb} and, (\textit{iii}) regarding tree-level graviton scattering amplitudes that cannot exist in four dimensions in case one considers a nontrivial four-dimensional limit of the EGB theory~\cite{Bonifacio:2020vbk}.} The $4D$ EGB theory may be invalid under these mentioned concerns. However, as viable alternatives to the classical {Einstein theory of gravity}, a lot of work has been done in understanding the nature of new EGB theory in this framework~\cite[see,
e.g.][]{Liu20egb,Guo20egb,Wei20egb,Kumar20egb,Konoplya20egb,Churilova20egb,Malafarina20egb,Aragon20egb,Mansoori20egb,Ge20egb,Rayimbaev20egb,Chakraborty20egb,Odintsov20egb,Odintsov20plb,Lin20egb,Aoki20egb,Shaymatov20egb,Islam20egb,Singh20-egb}. 

{In the recent literature there are numerous studies involving the spinning particle motion~\cite{Zhang20egb}, the analysis of GB term that affects on the superradiance~\cite{Zhang20aegb},  analysis of testing the strong and weak cosmic censorship conjectures~\cite{Mishra20:GReGr,Yang20b}, particle motion and plasma effect on weak gravitational lensing~\cite{Atamurotov21JCAP}, the stability of linearized equations of motion~\cite{Aguilar19} and Bondi-Hoyle accretion around $4D$ EGB black hole~\cite{Donmez2021egb}.  Later on, the charged and rotating solution analogues were obtained in $4D$ EGB theory (see for example \cite{Fernandes20plb,Kumar20egb}). Following to \cite{Glavan20prl}, 3D BTZ black hole solution in new EGB theory was obtained in Ref.~\cite{Hennigar20PLB}. Recently, particle motion and weak gravitational lensing have been studied around 3D BTZ black hole in novel EGB gravity~\cite{Narzilloev21BTZ}. Addition interesting aspect related to  studying  properties of GB black hole in higher $D$ dimensions was also developed in Refs.~\cite{Shaymatov20-pl,Dadhich22a,Wu21egb}.}   

The spacetime singularities which arise due to the gravitational collapse at the end state of stellar evolution should always been hidden behind the event horizon. If the event horizon disappears then naked singularity will destroy the spacetime and the law of causality. To solve this issue, Penrose proposed a conjecture known as weak cosmic censorship conjecture (WCCC) in $1969$ \cite{Penrose69}. It states that the singularity which is collapsed due to the gravitation must be hidden behind the event horizon and cannot be seen by the external observer at far distance. Obviously without the horizon, any particle that is emitted from the naked singularity would reach an outside observer and causes the disruption in the spacetime. Schwarzschild black hole is the only black hole where this disruption does not occur because the horizon never disappears. Anyhow, there are black holes which obey the extremal conditon under which no horizon of a black hole exists. If the naked singularity disappears due to the extremal condition, then the WCCC becomes invalid. Since under the extremal condition the horizon prevents the measurement of singularity, therefore, the main focus of WCCC is that if a black hole can violate the extremal condition through overcharging or overspinning. Though a general proof is still left but it has become the foundation of the black hole physics. Many particular attempts were made to prove this conjecture.

{Wald was the first one who proposed a gedanken experiment in $1969$ to check its validity by throwing a test particle with a sufficient amount of charge and angular momentum. He threw a test particle in extremal Kerr-Newman black hole and found that WCCC will never be destroyed under the first order perturbation \cite{Wald74b}. Later on, Hubeny  
showed that the horizon of a near extremal charged black hole can be destroyed by the charged particle \cite{Hubeny99}. The same result is obtained for near extremal Kerr and Kerr-Newman black holes by Jacobson and Sotiriou \cite{Jacobson09,Saa11}. There are many studies which reveals the relationship between the overcharging/overspinning and cosmic censorship conjecture \cite{deFelice01,Bouhmadi-Lopez10,Matsas07PRL,Richartz08,Hod08PLB,Matsas09,Richartz11,Semiz15,Rocha14,Shaymatov15,Song18,Duztas18,Jana18,Duztas-Jamil18b,Jiang20,Yang20a}. It was also shown that overcharging/overspinning of a black hole is possible, i.e.  BTZ black holes \cite{Rocha11,Duztas-Jamil20} and Higher dimensional Myers-Perry rotating black hole \cite{Bouhmadi-Lopez10}. Further, charged scalar and test fields were also considered in testing the  WCCC~\cite{Toth12,Duztas13,Duztas14,Duztas16,Duztas21}. Later, Gwak explored both extremal and non-extremal Kerr anti de-Sitter black holes and proved that they cannot be overspun by the scattering process in an infinitesimal time interval \cite{Gwak18a}.} 

{Also an extensive analysis involving scalar test fields around black holes has been done on these lines \cite{Gwak19JCAP,Chen20,Gwak20} testing the validity of the WCCC. There is a large amount work that shows that five dimensional rotating black holes \cite{Shaymatov19a}, BTZ black hole \cite{Zeng19-BTZ}, higher dimensional rotating/charged AdS black holes \cite{Chen19,Gwak19JCAP,Shaymatov21d} cannot be overspun/overcharged, and hence WCCC is always preserved. In all above mentioned studies, back-reaction effects are ignored. However, these effects are involved in following studies~\cite{Hod08PRL,Barausse10,Barausse11,Colleoni15a,Colleoni15b,Isoyama11,Shaymatov19b}. Recently, the new gedanken experiment has been developed by Sorce and Wald \cite{Sorce-Wald17,Wald18} addressing to consider the nonlinear order perturbations. They showed that a black hole obeys the WCCC under nonlinear order perturbations.   There are several investigations that addressed overcharging/spinning of the black holes in different settings under nonlinear order perturbations~\cite[see, e.g.][]{An18,Ge18,Ning19,Yan-Li19,Jiang20plb,Shaymatov19c,Shaymatov20a}. }

The WCCC has a strong connection with the laws of thermodynamics. If we study the gedanken experiment, we can better understand the relationship. Recently, Gwak and many other researches of WCCC in extended phase space have gained a lot of attention but the point of controversy is that when a particle having energy $E$ is thrown inside a black hole then instead of enthalpy internal energy of the black hole increases \cite{Gwak17JHEP}. The important flaw of this point of view is that it violates the second law of black hole thermodynamics and so it was opposed by many researchers. The fact is that, when a particle is thrown into a black hole, it is basically enthalpy which increases due to the particle's energy rather than the internal energy \cite{Hu19}. Hence, this argument preserves the second law of thermodynamics.

In this paper, we investigate the WCCC for the charged EGB BTZ black hole by the help of scattering of a massive charged scalar field and a charged test particle in the normal phase space, respectively. For scattering of a charged scalar field, our result shows that the opposite is the case as compered to the one for a black hole in $4D$ EGB gravity \cite{Yang20b}, i.e., the black hole can be overcharged and the WCCC is not valid. For a charged test particle we show that both extremal and near-extremal EGB BTZ black holes can be overcharged. { 
In this context, we wish to point out that our study in this field may add to a vast literature of theory that helps us to understand qualitatively how EGB theory behaves and exhibits striking departures of the geometry in Einstein theory of gravity and it may prove useful to learn better its remarkable gravitational and thermodynamical properties. } 

The paper is organized as follows. In Sec.~\ref{Sec:BTZ} we briefly discuss the basic properties and the thermodynamics of the charged EGB BTZ black hole. In Sec.~\ref{Sec:scalar} we consider the charged massive scalar field and build up the background for studying the WCCC with the help of this charged massive scalar field in Sec.~\ref{Sec:sf_wccc}. In Sec.~\ref{Sec:tp_wccc} we discuss the WCCC with the help of a charged test particle. We conclude our study with the discussion of the main results in Sec~\ref{Sec:conclusion}. Here, the metric signature is $(-,+,+)$ with $c=G=1$.

\section{The charged 3D BTZ black hole in EGB gravity}\label{Sec:BTZ}

The static solution of BTZ black hole in EGB gravity can easily be generalized by adding the non-linear electro-magnetic field which is given by the following action:
{ 
\begin{eqnarray}\label{1}
\mathcal{S}&=&\frac{1}{16\pi}\int\sqrt{-g} d^{3}x\Big\{R-2\Lambda+\alpha \Big[\phi L_{GB}\nonumber\\&&+ 4G^{\mu\nu}\partial_{\mu}\phi\partial_{\nu}\phi-4\left(\partial\phi\right)^2\square\phi\nonumber\\&&+2\big(\left(\triangledown\phi\right)^2\big)^2 \Big]+{L}(F)\Big\}\, ,
\end{eqnarray}
with $\alpha$ being the GB coupling parameter, $\phi$ is the scalar field and $F=F_{\mu\nu}F^{\mu\nu}$. Note that the GB term, $L_{GB}$, is given by
\begin{eqnarray}
L_{GB}=R_{\mu\nu\lambda\delta}R^{\mu\nu\lambda
	\delta} - 4R_{\mu\nu}R^{\mu\nu}+R^2\, ,
\end{eqnarray}
where $R$ refers to the scalar curvature. If one considers the Maxwell linear theory of electromagnetic field  ${L}_{M}=-F$ and the theory with $\phi=\ln({r}/{l})$, where $l$ is an integration constant, it is then possible to obtain}
\begin{eqnarray}\label{2}
ds^{2}&=&-f_{GB}dt^{2}+\frac{dr^{2}}{f_{GB}}+r^{2}d\varphi^{2},
\end{eqnarray}
where 
\begin{eqnarray}\label{Eq:mf}
f^{\pm}_{GB}(r)&=&-\frac{r^{2}}{2\alpha}\Big(1\pm\sqrt{1+\frac{4\alpha}{r^2}f_{E}}\Big)\, , \label{3} 
\end{eqnarray}
where the function $f_{E}$ and vector potential $A$ can be defined by 
\begin{eqnarray}
f_{E}&=&\frac{r^{2}}{l^2}-M-2Q^{2}\ln\Big(\frac{r}{r_{0}}\Big),\label{5}\\
A&=&-Q\ln\Big(\frac{r}{r_0}\Big) dt\, ,\label{6}
\end{eqnarray}
where $r_0$, $M$ and $l$ are arbitrary integration constants. Physically, the quantity $r_0$ corresponds to an arbitrary reference point for the energy of the black hole, $M$ corresponds to the Arnowitt-Deser-Misner (ADM) mass and $l$ is the length scale such that $\Lambda=-1/l^{2}$.  {Here, it is worth noting that we focus on the 'negative' branch of solution because it reduces to standard BTZ solution in Einstein gravity in the limit of $\alpha\ll 1$, and thus Eq.~(\ref{Eq:mf}) can be written as~\cite{Hennigar20PLB}:  
\begin{eqnarray}
\lim_{\alpha\rightarrow 0} f(r) &=& \frac{r^2}{l^2}-M-2Q^{2}\ln\Big(\frac{r}{r_{0}}\Big)\nonumber\\&-&\frac{\alpha}{r^2}\left(\frac{r^2}{l^2}-M-2Q^{2}\ln\Big(\frac{r}{r_{0}}\Big)\right)^2+O(\alpha^2)\, .
\end{eqnarray}
The above equation is exactly similar to the charged BTZ solution in the Einstein gravity in the limiting case of small $\alpha$. Thus, we further restrict ourselves to $f^{-}_{GB}(r)$ that describes the charged BTZ black hole in the novel EGB gravity.  
The metric function (i.e. $f^{-}_{GB}(r)$) can be then written in the following form
\begin{eqnarray}\label{7}
f(r)=-\frac{r^{2}}{2\alpha}\frac{\Big[1-\sqrt{\left(1+\frac{4\alpha}{r^2}f_{E}\right)^2}\,\Big]}{1+\sqrt{1+\frac{4\alpha}{r^2}f_{E}}}=\frac{2f_{E}}{1+\sqrt{1+\frac{4\alpha}{r^2} f_{E}}}\, .
\end{eqnarray}}
The event horizon is determined by the equation $f(r)=0$ which is equivalent to $f_{E}=0$.  
For the extremal charged EGB black hole it is determined by $\frac{\partial M}{\partial r_{+}}$ \footnote{{We find $M$ by setting metric function equals to zero and solving it with respect to $M$}}, which gives the degenerate horizon as
\begin{eqnarray}\label{8}
r_{\pm}&=&\pm Ql\, .
\end{eqnarray}
Here, $r_{+}$ denotes the event horizon of  black hole.

{The thermodynamics of a black hole in $(2+1)$ dimensions is similar to the one for $(3+1)$ dimensions. It can be expected that $f_{GB}(r)$ remains valid for the charged EGB BTZ black holes in any theory of non-linear electrodynamics. Thus, the temperature of the black hole can be calculated as \cite{Hennigar21egb}
\begin{eqnarray}\label{9}
T&=&\frac{f^{\prime}_{E}(r_{+})}{4\pi}=\frac{f^{\prime}_{GB}(r_{+})}{4\pi}\, ,
\end{eqnarray}
whereas, the area of the event horizon and the entropy are given by
\begin{eqnarray}\label{10}
A&=&2\pi r_{+}\, \mbox{~~and~~} S=\frac{1}{2}\pi r_{+}\, . 
\end{eqnarray}
The electric potential is given by
\begin{eqnarray}\label{12}
\Phi&=&Q\ln\Big(\frac{r}{r_0}\Big)\, .
\end{eqnarray}
The standard first law of thermodynamics of the black hole \cite{Hennigar21egb} is written as 
\begin{eqnarray}\label{13}
\delta M&=&T\delta S+\Phi\delta Q+V\delta P+\Pi_{\alpha}\delta \alpha\, .
\end{eqnarray}
The left hand side is the change in  mass, while the first and the second  terms on the right hand side represent changes in energy due to entropy and electromagnetism. The third term includes the variations of the cosmological constant. In this case we assume it is just constant. However, it is taken $\Pi_{\alpha}=0$ since thermodynamic parameters are the same for any value of GB coupling parameter $\alpha$, thus leading to similar forms as compared to the one for BTZ black hole in Einstein theory of gravity.  Consequently, the first law of thermodynamics is a statement of energy conservation. }

\section{Charged massive scalar field in charged EGB spacetime in $2+1$ dimensions}\label{Sec:scalar}

We study the scattering of charged massive scalar field in the $2+1$ charged EGB spacetime background. The charged massive scalar field $\Psi$ with mass $\mu_0$ and charge $q$ minimally coupled to the gravity is governed by the equation
\begin{eqnarray}\label{18}
\Big[(\nabla^{\nu}-iqA^{\nu})(\nabla_{\nu}-\iota qA_{\nu})-\mu_0^{2}\Big]\Psi&=&0.
\end{eqnarray}
This can be expanded as
\begin{eqnarray}\label{19}
&&\frac{1}{\sqrt{-g}}\partial_{\mu}\Big[\sqrt{-g}\,g^{\mu\nu}\partial_{\nu}\Psi\Big]-\frac{iq}{\sqrt{-g}}\partial_{\mu}\Big[\sqrt{-g}\,g^{\mu\nu} \left(A_{\nu}\Psi\right)\Big]\nonumber\\&&-\frac{iqA_{\mu}}{\sqrt{-g}}\Big[\sqrt{-g}\,g^{\mu\nu} \partial_{\nu}\Psi\Big]-q^{2}g^{\mu\nu} A_{\mu}A_{\nu}\Psi-\mu_0^{2}\Psi=0\, ,\nonumber\\
\end{eqnarray}
which on further simplification gives
\begin{eqnarray}\label{20}
&&\frac{1}{r}\Big[\partial_{t}\Big(-\frac{r}{f(r)}\partial_{t}\Psi\Big)+\partial_{r}\Big(rf(r)\partial_{r}\Psi\Big)+\partial_{\phi}^2\Psi\Big]-\mu_0^{2}\Psi\nonumber\\&&-\frac{2iqQ}{f(r)}\ln\Big(\frac{r}{r_{0}}\Big)\partial_{t}\Psi+\frac{q^{2}Q^{2}}{f(r)}\ln^{2}\Big(\frac{r}{r_{0}}\Big)\Psi=0\ . 
\end{eqnarray}
Since the spacetime is static and spherically symmetric, here the complex scalar field can be decomposed by assuming the ansatz
\begin{eqnarray}\label{21}
\Psi(t,r,\varphi)&=&e^{-i\omega t}R_{lm}(r)Y_{lm}(\varphi)\, ,
\end{eqnarray}
where $R_{lm}(r)$ are the radial functions and $Y_{lm}(\varphi)$ are the spherical harmonic functions. Using this decomposition in Eq. (\ref{20}) we get the radial and angular parts for the equation of motion as
\begin{eqnarray}
&&\frac{1}{r}\frac{\partial}{\partial r}\Big(rf(r)\frac{\partial R_{lm}}{\partial r}\Big)\nonumber\\&&+\left(\frac{\Big[\omega-{qQ}\ln\Big(\frac{r}{r_{0}}\Big)\Big]^{2}}{f(r)}-\frac{l(l+1)}{r^2}-\mu_0^2\right)R_{lm}=0,\label{22}\\
&&\frac{\partial^{2}Y_{lm}}{\partial\varphi^{2}}=-l(l+1)Y_{lm},\label{23}
\end{eqnarray}
where $l(l+1)$ is the separation constant such that $l>0$. Now the solution to Eq. (\ref{23}) is expressed through the spherical harmonic functions and we are more interested in the radial part. For finding solution to the radial part we introduce the tortoise coordinate as 
\begin{eqnarray}\label{24}
r_{\ast}&=&\int\frac{dr}{f(r)}\, ,
\end{eqnarray}
then the radial equation takes the form
\begin{eqnarray}\label{25}
&&\frac{d^{2}R_{lm}}{dr_{\ast}^{2}}+\frac{f(r)}{r}\frac{dR_{lm}}{dr_{\ast}}+\Big[\omega-{qQ}\ln\Big(\frac{r}{r_{0}}\Big)\Big]^{2}R_{lm}\nonumber\\&&-f(r)\Big[\frac{l(l+1)}{r^2}+\mu_0^2\Big]R_{lm}=0\, .
\end{eqnarray}
At the horizon, $r=r_{+}$, the function $f(r)$ vanishes and the above equation reduces to 
\begin{eqnarray}\label{26}
\frac{d^{2}R_{lm}}{dr_{\ast}^{2}}+\Big[\omega-{q}{Q}\ln\Big(\frac{r_+}{r_{0}}\Big)\Big]^{2}R_{lm}=0\, ,
\end{eqnarray}
{and the solution to this equation is given by}
\begin{eqnarray}\label{27}
R_{lm}\sim exp\Big[\pm i\Big[\omega-{qQ}\ln\Big(\frac{r_+}{r_{0}}\Big)\Big]r_{\ast}\Big]\, .
\end{eqnarray}
Note that we choose the negative solution since it corresponds to the falling of wave modes {with frequency $\omega$} into the black hole. Hence, one can write the expression foe the charged scalar field in the following form
\begin{eqnarray}
 \Psi&=&e^{-i\omega t}e^{- i\Big[\omega-{qQ}\ln\Big(\frac{r_+}{r_{0}}\Big)\Big]r_{\ast}}Y_{lm}(\varphi)\, .
\end{eqnarray}

Taking into consideration the above wave function we further test the validity of the WCCC in the case of charged scalar field interacting with the charged BTZ EGB black hole in $2+1$ dimensions. For the final state of the black hole parameters we use the energy and electric charge fluxes of the scalar field.  This is what we wish to address in the next section.

\section{Overcharging black hole with charged massive scalar field}\label{Sec:sf_wccc}

The metric function $f_{E}$ determines the event horizon of black hole  i.e.
\begin{eqnarray}\label{14}
\Delta&\equiv&f_{E}=\frac{r^2}{l^2}-M-2Q^{2}\ln\Big(\frac{r}{r_0}\Big).
\end{eqnarray}
The metric function $\Delta$ takes the minimal value at the point $r=Ql(1+\lambda)$ for a near-extremal black hole so it gives
\begin{eqnarray}\label{15}
\Delta_{\rm{in}}&=& M- {Q}^2\left[ 1- 2\ln\Big( \frac{l\,Q}{r_0}\Big) \right]= 2 Q^2 \lambda^2\, .
\end{eqnarray}
It is certain that {$\lambda=0$} corresponds to an extremal case. We give further details for Eq.~(\ref{15}) in the next section. In fact after particle absorption, the black hole parameters change in the way that
\begin{eqnarray}\label{16}
M&\rightarrow&M_{fin}=M+dE\, ,\\
Q&\rightarrow&Q_{fin}=Q+dQ\, .
\end{eqnarray}
Hence, $\Delta_{in}$ becomes $\Delta_{fin}$ i.e.
\begin{eqnarray}\label{17}
\Delta_{fin}&=&\Delta_{fin}\Big(M+dE,Q+dQ\Big)=\nonumber\\
&=&\Delta_{in}+\Big(\frac{\partial\Delta_{in}}{\partial M}\Big)dE+\Big(\frac{\Delta_{in}}{\partial Q}\Big)dQ=\nonumber\\
&=&2Q^{2}\lambda^2+\delta E-2Q\ln \left(\frac{Q l}{r_0}\right)^2dQ\, .
\end{eqnarray}
From the above equation, if and only if $\Delta_{fin}<0$ is the case then it indicates that a near-extremal black hole can be overcharged under the perturbation of charged massive scalar field. If we include the solution with $\left({Q l}/{r_0}\right)^2>1$, then the function $\ln \left({Q l}/{r_0}\right)^2$ always increases, leading to the negative result, $\Delta_{fin}<0$, even in the extremal case $\lambda=0$. However, we may approach to this issue from different prospective. For a near extremal black hole, radius of the event horizon, $r_{+}$, is extremely close to the minimum value $r_{min}$, and we define an infinitesimal distance $\lambda$ between the event horizon, $r_{+}$, and the minimal point $r_{min}$ as $r_{+}=r_{min}+Ql\lambda$, where $\lambda=0$ is responsible for an extremal case and $\lambda>0$ for a near-extremal case. Let us then define the final state of the black hole parameters. If a massive charged scalar field scatters off the black hole it would result in a black hole’s mass and charge, i.e. $M +\delta E$ and $Q +\delta Q$. Here $\delta E$ and $\delta Q$, respectively, refer to the changes in mass and charge of the black hole and can be calculated from the energy and electric charge fluxes of the scalar field under the scattering. 

It is worth noting that we consider a wave mode, $m=0$, associated with the angular momentum due to the fact that the $3D$ charged black hole considered here has no rotation. Hence, we focus on a single wave mode, $l\neq 0$, through the calculations. For the charged scalar field the energy-momentum tensor is defined by 
\begin{eqnarray}\label{Eq:en-mom}
 T^\mu_\nu &=&\frac{1}{2}\mathcal{D}^\mu\Psi\partial_\nu\Psi^*+ \frac{1}{2}\mathcal{D}^{*\mu}\Psi^*\partial_\nu\Psi \nonumber\\ &-&\delta^\mu_\nu\left(\frac{1}{2} \mathcal{D}_\alpha\Psi\mathcal{D}^{*\alpha}\Psi^*- \frac{1}{2}\mu_{0}\Psi\Psi^*\right)\, ,
\end{eqnarray}
where $\mathcal{D}$ is given by
\begin{equation}
  \mathcal{D}=\partial_\mu-iqA_\mu\, .
\end{equation}
Then the energy and charge fluxes at the horizon can be calculated as 
\begin{eqnarray}\label{Eq:En-flux}
   \frac{dE}{dt}&=&\int T^r_t\sqrt{-g} \, d\phi=\omega(\omega-q\Phi_{+})r_{+}\, ,\\
   \frac{dQ}{dt}&=&-\int j^r\sqrt{-g} d\phi=q(\omega-q\Phi_{+})r_{+}\, ,
\label{Eq:Q-flux}
\end{eqnarray}
where the electric current for the charged scalar field is
given by 
\begin{eqnarray}\label{ecurrent}
  j^\mu=-\frac{1}{2}iq(\Psi^{*}\mathcal{D}^{\mu}\Psi-\Psi \mathcal{D}^{*\mu}\Psi^{*}).
\end{eqnarray}
In doing so, the change in the energy and charge of the black hole can be defined by the change in those parameters of the falling in/ingoing charged scalar field; Eqs.~(\ref{Eq:En-flux}-\ref{Eq:Q-flux}).

Now if we consider infinitesimal time interval $dt$, the changes in mass and charge of the black hole are given by
\begin{eqnarray}\label{Eq:dE}
   dE &=& \omega(\omega-q\Phi_{+})r_{+} \, dt\, , \label{Echange} \\
   dQ &=&  q(\omega-q\Phi_{+})r_{+}\,dt\, . 
   \label{Eq:dQ}
\end{eqnarray}
From the equations in the above, the scalar field flowing into the black hole with wave modes, $\omega > q {\Phi_\text{+}}$, would carry the energy and charge to its mass and charge. If the scalar field is defined by the wave modes through $\omega < q {\Phi_\text{+}}$, and then the energy $\delta E$ and charge $\delta Q$ fluxes correspond to the outging charged scalar field with the energy and charge extracted from the black hole, i.e. it defines the super-radiance $\omega < \omega_s$~\cite{Duztas16}.

Using Eqs.~(\ref{Eq:dE}-\ref{Eq:dQ}) along with Eq.~(\ref{17}) we get
\begin{eqnarray}\label{30}
\Delta_{fin}&=&2Q^2\lambda^2+q^{2}\Big(\frac{\omega}{q}-Q\ln \frac{r_+}{r_0}\Big)\nonumber\\&\times &\Big[\frac{\omega}{q}-4Q\ln \left(\frac{lQ}{r_0}\right)\Big]r_{+}dt\, .
\end{eqnarray}
Let us then first consider an extremal case, i.e. $\lambda=0$. After the absorption of the charged scalar field the above equation takes the following form 
\begin{eqnarray}\label{31}
\Delta_{fin}&=&\left[\frac{\omega}{q}-Q\ln \left(\frac{lQ}{r_0}\right)\right]\left[\frac{\omega}{q}-4Q\ln \left(\frac{lQ}{r_0}\right)\right]q^{2}r_{+}dt\, .\nonumber\\
\end{eqnarray}
It is certain that the charged scalar field with mode must satisfy the following condition
\begin{eqnarray}\label{32}
Q\ln \left(\frac{lQ}{r_0}\right)<\frac{\omega}{q}<4Q\ln \left(\frac{lQ}{r_0}\right)\, ,
\end{eqnarray}
so that an extremal black hole is overcharged. If the condition given in Eq.~(\ref{32}) for the charged scalar field is the case then the result of $\Delta_{fin}$ is always negative, in turn indicating that an extremal black hole can be overcharged, i.e. the inevitable violation of the WCCC occurs.  

Now we turn to the near-extremal case, i.e. $\lambda\neq 0$. Let us take the charged scalar field with the following mode for a near-extremal black hole 
\begin{eqnarray}
 \frac{\omega}{q}=\frac{1}{2}Q\ln \frac{r_+}{r_0}+2Q\ln \left(\frac{lQ}{r_0}\right)\, .
\end{eqnarray}
We then check that whether the validity of the WCCC is violated after absorption by the black hole of  the charged scalar field with the selected  mode. Thus, after the absorption of the charged scalar field, the minimal value of the function $\Delta_{fin}$ can be approximated as 
\begin{eqnarray}\label{Eq:fin_near}
\Delta_{fin}&\approx&2Q^2\lambda^2-3q^{2}\left[Q\ln \left(\frac{lQ}{r_0}\right)\right]^2 l\,Q \lambda\, .
\end{eqnarray}
We assume that the absorption of the scalar field with the energy and charge occurs in an infinitesimally small time interval with $dt\sim \lambda$. From Eq.~(\ref{Eq:fin_near}), $\Delta_{fin}$ can give negative value since $l\,Q\gg \lambda$ always. Hence, it is evident that the near-extremal 3D charged EGB black hole can be overcharged by the scattering of the charged scalar field, and consequently the WCCC no longer holds.  

From the above results we have shown that it is possible for the charged scalar field to overcharge both an extremal and near-extremal 3D charged BTZ black holes in EGB gravity, accordingly resulting in violating the WCCC. We have noticed that the violation of the WCCC does not depend on the GB coupling parameter $\alpha$. Unlike a four-dimensional charged EGB black hole, it can be overcharged, and consequently it is unstable under the scalar field perturbation. It is a remarkable property of a charged BTZ EGB black hole in $(2+1)$ dimensions.

\section{Overcharging black hole in EGB gravity with charged massive particle}\label{Sec:tp_wccc}

\subsection{Overcharging  extremal BTZ black hole in EGB gravity}

Here we then study a massive charged particle absorption by an extremal BTZ black hole with the total mass $M$ and charge $Q$ with the aim to clarify whether the WCCC can be violated for an extremal/nearly extremal black holes in EGB gravity.

Note that the required condition for existence of black hole solution $f_{E}(r_{\rm{min}})\leq0$ must be held, i.e.
\begin{equation}\label{Eq:extremal}
\Delta\equiv M- {Q}^2\left[ 1- 2\ln\Big( \frac{l\,Q}{r_0}\Big) \right] \geq 0\, .
\end{equation}
From the above equation the following function ${Q}^2\left[ 1- 2\ln\left( {l\,Q}/{r_0}\right) \right]$ no longer exists at $Q=0$ and ${l\,Q}/{r_0}=e^{1/2}$. However it reaches its maximum value for ${l\,Q}/{r_0}=1$. In this regard, $\Delta$ takes values that are always larger than zero in the case when $M>1$, and accordingly an ordinary black hole exists with inner and outer horizons, i.e. $r_{-}$ and $r_{+}$. 

First we investigate an extremal black hole for which $\Delta_{\rm{in}}=0$ is always satisfied. Once test particles with energy $\delta E$ and charge $\delta Q$ are absorbed, the black hole is perturbed and then its final state is given by $M+\delta E$ and $Q+\delta Q$, respectively. Then we can explore  whether it is possible to overcharge extremal black hole in the cases when $M<1$. For the cases when $M<1$ and $(lQ/r_0)^2<1$, it is likely to have $r_+<Ql$, thereby leading to the minimum energy which would be negative. Test particle with energy required for overcharging must reach the black hole horizon in the way  that the final state of the black hole parameters gives rise to a naked singularity spacetime. This represents the  maximum energy of the particle which is obtained from the following condition 
\begin{eqnarray}\label{Eq:fin_ex}
 \Delta_{\rm{fin}}&=&(M+\delta E)-\left( {Q+\delta Q}\right)^2 \nonumber\\&+& 2\left( {Q+\delta Q}\right)^2\ln\left[ \frac{l\left(Q+\delta Q\right)}{r_0} \right]<0\, . \label{deltafin}
\end{eqnarray}
It has to be noted here  that we assume test particle approximation, i.e.  $\delta E\ll M$ and $\delta Q\ll Q$. Thus, in this approximation we choose $\delta Q=\epsilon Q$ with small $\epsilon \ll 1$. In doing this the test particle approximation holds well, thus allowing to make expansion in series for small $\epsilon$, for example  
$\ln(1+\epsilon)\simeq\epsilon - \frac{\epsilon^2}{2}+O(\epsilon^3)$. According to  Eq.~\ref{Eq:fin_ex}, the last term can be expanded up to second order in $\epsilon$
\[
\ln\left[ \frac{l\left( Q+\delta Q\right)}{r_0} \right]=\ln\left[ \left( \frac{l Q}{r_0}\right)\right]+\epsilon-\frac{\epsilon^2}{2}.
\]
Then Eq.~(\ref{deltafin}) yields as
\begin{eqnarray}
&&M +\delta E- {Q}^2-\epsilon^2 {Q}^2 -2\epsilon  {Q}^2 + 2\left\{  {Q}^2+\epsilon^2  {Q}^2+2\epsilon {Q}^2\right\}
\nonumber\\&&\times \left\{\ln\left[ \left( \frac{l\,Q}{r_0}\right) \right] +\epsilon -\frac{\epsilon^2}{2} \right\}<0 .
\label{deltafin1}
\end{eqnarray}
From straightforward calculations on the basis of $\Delta_{\rm{in}}=0$, we have 
\begin{equation}
\delta E<\delta E_{\rm{max}}=-2(\epsilon^2+2\epsilon){Q}^2\ln\left[\left( \frac{l\,Q}{r_0}\right)\right]-2\epsilon^2  {Q}^2\, . \label{deltafin2}
\end{equation}
To keep $\delta E>0$ in the above equation the following condition $Q^2<\frac{r_0^2}{l^2}e^{(-2\epsilon)/(\epsilon +2)}$ is always satisfied and gives the appropriate range for black hole charge. Then the black hole can have the total electric charge up to $Q^2<0.990099$. For further analysis for simplicity of calculations we assume that  $l=r_0$. Then we choose test particle charge $\delta Q=\epsilon Q$ and it's energy to be in the range $0<\delta E<  \delta E_{\rm{max}}$. With these assumptions and choices we test whether an extremal black hole is overcharged. 
 
As stated earlier $\delta E_{\rm{min}}$ is negative for the cases we considered here. For given $Q$ we choose $\delta E$ from the range mentioned above. For the above approach to overcharge the extremal black hole with $(Q)^2=0.5$ we now start to explore the process numerically. We can have $M=0.846574$ by imposing the condition $\Delta_{\rm{in}}=0$. Keeping in mind $\delta Q=\epsilon Q$ with $\epsilon=0.01$, we obtain $E_{\rm{max}}=0.00687$. It is worth noting that $\delta E_{\rm{max}}\lesssim M\epsilon$ is small enough and  satisfies test particle approximation. We then evaluate $\Delta_{\rm{fin}}$ for chosen $\delta E=0.005<\delta E_{\rm{max}}$ for test particle. It is given by 
\begin{eqnarray} \label{Eq:ex_res}
 \Delta_{\rm{fin}}&=&M+\delta E-\left( {Q+\delta Q}\right)^2 +2\left( {Q+\delta Q}\right)^2\nonumber\\&&\times \ln\left[ \frac{l\left(Q+\delta Q\right)}{r_0} \right]=-0.00186539\, .
\label{exext}
\end{eqnarray}
The above result indicates that the charged particle having suitable parameters can cross the black hole horizon without any restriction as there exists no turning point. Therefore, the extremal BTZ EGB black hole can be overcharged and can be turned into a naked singularity, resulting in violating the WCCC.

\subsection{Overcharging near-extremal BTZ black hole in EGB gravity}

In the above it has been demonstrated that an extremal BTZ EGB black hole can be overcharged. Let us now consider a near-extremal BTZ black hole. As mentioned earlier the horizon is given by $r_+=Ql$ for an extremal black hole. For a near-extremal black hole one can define its horizon as 
\begin{equation}
r_+={Q l}(1+\lambda)\, ,
\end{equation}
with the parameter $\lambda$ which respectively represents a near extremality. For our purpose we choose $\lambda$ being smaller as that of unity so as to describe a near-extreaml black hole. In the limit of $\lambda=0$ it corresponds to an extremal black hole. Taking into consideration  $\lambda$  Eq.~(\ref{Eq:extremal}) is defined by  
\[
-2 Q^2 \ln \left(\frac{l Q (1+\lambda)}{r_0}\right)-M+Q^2 (1+\lambda)^2=0\, ,
\]
and then we have 
\begin{eqnarray}
\Delta_{\rm{in}}&=& M- {Q}^2\left[ 1- 2\ln\Big( \frac{l\,Q}{r_0}\Big) \right]= 2\lambda^2 Q^2\, . \label{deltain}
\end{eqnarray}
For a near-extremal black hole the condition  $\Delta_{\rm{in}}=2\lambda^2 Q^2$ is always satisfied as shown in the above equation. As done above, in order to overcharge the near-extremal black hole the condition $\Delta_{\rm{in}}<0$ has to be satisfied. We approach to  this issue in the similar way as done for an extremal black hole case. Assuming $\delta Q=Q\epsilon$ we obtain $\delta E_{\rm{max}}$ in the case of near-extremal black hole. Using Eqs.~(\ref{deltain}) and (\ref{deltafin1}) we find $\delta E_{\rm{max}}$ as 
\begin{eqnarray}\label{deltafin3}
\delta E_{\rm{max}}&=&-2(\epsilon^2+2\epsilon){Q}^2\ln\left[\left( \frac{l\,Q}{r_0}\right)\right]-\left[2\epsilon^2+2\lambda^2\right]{Q}^2\, .\nonumber\\
\end{eqnarray}
Then we explore it numerically. Setting $Q^2=0.5$ and $\epsilon=\lambda=0.01$ together with  $\Delta_{\rm{in}}=2\lambda^2Q^2$, one can easily obtain the black hole mass as $M=0.8466$ which, in turn, leads to $\delta E_{\rm{max}}=0.00676613$ based on the  Eq.~(\ref{deltafin3}). We are then able to choose $\delta E=0.005$ in the range $E_{\rm{min}}<E<\delta E_{\rm{max}}$, where $\delta E_{\rm{min}}= Q \delta Q \ln ({r_+}/{r_0})<0$ always. Eq.~(\ref{Eq:ex_res}) for near-extremal black hole takes the following numerical value
\begin{eqnarray}
 \Delta_{\rm{fin}}&=&M+\delta E-\left( {Q+\delta Q}\right)^2 +2\left( {Q+\delta Q}\right)^2\nonumber\\&&\times \ln\left[ \frac{l\left(Q+\delta Q\right)}{r_0} \right]=-0.00176539\, .
\label{exnearext}
\end{eqnarray}
The above result shows that $\Delta_{\rm{fin}}$ is negative and consequently near-extremal charged BTZ EGB black hole can also be overcharged, i.e. it is possible to violate the WCCC in EGB gravity, similarly to what is observed in $(2+1)$ dimensional black hole in Einstein gravity~\cite{Duztas-Jamil20}.\\

\section{Discussion and Conclusions}\label{Sec:conclusion}

In this paper, we have studied the validity of the WCCC for the charged BTZ black hole in (2+1)-dimensional EGB gravity with the help of the charged massive scalar field and charged test particle. We have studied properties of both the extremal and nearly extremal black holes. We have shown that it is possible for the charged scalar field to overcharge both an extremal and near-extremal black holes, thus resulting in violating the WCCC. Unlike a four-dimensional charged EGB black hole, interestingly, we have shown that it can be overcharged and thus the WCCC is not valid under the charged scalar field perturbation, resulting in not depending on the GB coupling parameter. The result would continue to do so for charged test particle perturbations. 

We have derived $\delta E_{\rm{min}}$ and $\delta E_{\rm{max}}$ energies for the overcharging test  particles. The test particles are not eligible to overcharge the black hole provided that $\delta E<\delta E_{\rm{min}}$ is fulfilled. 
If the particle's energy required for BH overcharging is  greater than $\delta E_{\rm{max}}$ it turns out that the test particle is not allowed to reach the black hole horizon. Hence, there exists no parameter space available for the particles that could turn nearly extremal black hole into over extremal one, thus no possibility for violating the WCCC.  

If $\delta E_{\rm{min}}<\delta E_{\rm{max}}$ is always satisfied then there exists parameter space available for charged test particles that could lead to overcharging of black hole. For the extremal BTZ black hole in (2+1)-dimensional EGB theory,  we have shown that the minimum energy that particle can have at the horizon becomes negative, and accordingly an extremal black hole can be overcharged. We have also shown that near extremal $(2+1)$ dimensional BTZ black hole in EGB theory can also be overcharged similar to the $(2+1)$ dimensional MTZ black hole in Einstein gravity.

\section*{Acknowledgements}

This work is partly supported by Grants F-FA-2021-432, and MRB-2021-527 of the Uzbekistan Ministry for Innovative Development and by the Abdus Salam International Centre for Theoretical Physics, Italy under the Grant No. OEA-NT-01.


\bibliographystyle{apsrev4-1}  
\bibliography{gravreferences,reference}

\end{document}